\documentclass[pra,showpacs,twocolumn,eqsecnum,]{revtex4}
\usepackage{graphicx}
\usepackage[urlcolor=blue, hyperindex, colorlinks, bookmarks=true,linkcolor=black,citecolor=black]{hyperref}
\begin{document}

\title{State and dynamical parameter estimation for open quantum systems}
\author{Jay Gambetta}
\affiliation{School of Science,
Griffith University, Brisbane 4111, Australia}
\author{H. M. Wiseman} \email{h.wiseman@gu.edu.au}
\affiliation{School of Science,
Griffith University, Brisbane 4111, Australia}

\begin{abstract}
Following the evolution of an open quantum system requires full knowledge of its dynamics. In this paper we consider open quantum systems for which the Hamiltonian is ``uncertain''. In particular, we treat in detail a simple system similar to that considered by Mabuchi [Quant. Semiclass. Opt. {\bf 8}, 1103 (1996)]: a radiatively damped atom driven by an unknown Rabi frequency $\Omega$ (as would occur for an atom at an unknown point in a standing light wave). By measuring the environment of the system, knowledge about the system state, and about the uncertain dynamical parameter, can be acquired. We find that these two sorts of knowledge acquisition (quantified by the  posterior distribution for $\Omega$, and the conditional purity of the system, respectively) are quite distinct processes, which are not strongly correlated. Also, the quality and quantity of knowledge gain depend strongly on the type of monitoring scheme. We compare five different detection schemes (direct, adaptive, homodyne of the $x$ quadrature, homodyne of the $y$ quadrature, and heterodyne) using four different measures of the knowledge gain (Shannon information about $\Omega$, variance in $\Omega$, long-time system purity, and short-time system purity).
\end{abstract}

 \maketitle

\newcommand{\beq}{\begin{equation}}
\newcommand{\eeq}{\end{equation}}
\newcommand{\bqa}{\begin{eqnarray}}
\newcommand{\eqa}{\end{eqnarray}}
\newcommand{\nn}{\nonumber}
\newcommand{\nl}{\nn \\ &&}
\newcommand{\dg}{^\dagger}
\newcommand{\rt}[1]{\sqrt{#1}\,}
\newcommand{\erf}[1]{Eq.~(\ref{#1})}
\newcommand{\Erf}[1]{Equation~(\ref{#1})}
\newcommand{\smallfrac}[2]{\mbox{$\frac{#1}{#2}$}}
\newcommand{\bra}[1]{\left\langle{#1}\right|}
\newcommand{\ket}[1]{\left|{#1}\right\rangle}
\newcommand{\ip}[2]{\left\langle{#1}\right|\left.{#2}\right\rangle}
\newcommand{\sch}{Schr\"odinger }
\newcommand{\schs}{Schr\"odinger's }
\newcommand{\hei}{Heisenberg }
\newcommand{\heis}{Heisenberg's }
\newcommand{\half}{\smallfrac{1}{2}}
\newcommand{\bl}{{\bigl(}}
\newcommand{\br}{{\bigr)}}
\newcommand{\ito}{It\^o }
\newcommand{\sq}[1]{\left[ {#1} \right]}
\newcommand{\cu}[1]{\left\{ {#1} \right\}}
\newcommand{\ro}[1]{\left( {#1} \right)}
\newcommand{\an}[1]{\left\langle{#1}\right\rangle}
\newcommand{\implies}{\Longrightarrow}
\newcommand{\tr}[1]{{\rm Tr}\sq{ {#1} }}
\newcommand{\st}[1]{\left| {#1} \right|}
\newcommand{\bfi}{{\bf I}_{[0,t)}}
\newcommand{\rhoc}{\rho_{{\bf I}}}
\newcommand{\singlecol}{\end{multicols}
                        \vspace{-0.5cm}

\noindent\rule{0.5\textwidth}{0.4pt}\rule{0.4pt}{\baselineskip}
                        \widetext }
\newcommand{\doublecol}{\noindent\hspace{0.5\textwidth}

\rule{0.4pt}{\baselineskip}\rule[\baselineskip]{0.5\textwidth}{0.4pt}
                        \vspace{-1cm}\begin{multicols}{2}\noindent}

\section{Introduction}

Quantum parameter estimation is a well-established area
\cite{Hel76,Hol82}, which is usually
formulated as follows. A known quantum state enters an apparatus that
performs an operation on the state. The operation, which is usually unitary
but need not be \cite{PoyCirZol97,ChuNie97},
is parameterized by one or more unknown
parameters. The goal is to estimate these parameters by making a
measurement on the (unknown) output state. Except in special cases, it is
not possible precisely to find out the unknown parameters from a
measurement
on a single system. Rather, the operation and measurement must be performed
repeatedly, on a sequence of identically prepared quantum systems.

There is a trivial sense in which it is possible to obtain complete
information about the unknown parameters from a single system. That is
by taking the output state after the measurement, and using it as the
next input state, having perhaps transformed it first. If the
transformation required is as difficult as preparing a new system from
scratch, then there is nothing to be gained by reusing the same
system. However, this scenario
of repeated measurements on a single system is useful pedagogically
to make the transition to continuously monitored systems with unknown
dynamical parameters. This transition is made by considering the limit
where the unknown
transformation is infinitesimally different from the identity,
and the repeat time is infinitesimal.

To the best of our knowledge, a theoretical treatment of
estimating an unknown dynamical
parameter by continuous observation of a system was first done by
Mabuchi \cite{Mab96}. His system was a two-level atom coupled to a
classically driven electromagnetic field mode in a cavity.
The unknown parameter was
the position of the atom. This is a dynamical parameter because it
determines the strength of the coupling between the atom and field
(the Rabi frequency).
The continuous monitoring considered was counting the photons that
escape through one of the cavity mirrors. Mabuchi used Bayesian statistics
to determine the posterior probability
distribution for the Rabi frequency. This represents the knowledge the
experimenter would have about the Rabi frequency given a particular
(typical) measurement
record. The measurement is continuous in time (monitoring) because in
any instant of time a photon may or may not be detected.

In this paper we are concerned with the same question, namely how
would an experimenter gain knowledge of an unknown dynamical
parameter from the measurement record resulting from monitoring the
system. We even choose a similar (but even simpler)
quantum system to that of Ref.~\cite{Mab96}, namely an atom driven by
a classical field of unknown Rabi frequency. However, our analysis
goes beyond, and has additional aims to, that of Ref.~\cite{Mab96}
(although we should note that extensions similar to the first three
outlined below were suggested in a footnote of that work.)

First, we consider the entire ensemble of possible measurement
records and parameter values, rather than just one (typical)
measurement record from one parameter value.

Second, we quantitatively characterize this ensemble by calculating
the average information gained (in bits) by the measurement,  as a
function of time.

Third, we consider different ensembles resulting from different
measurement schemes on the system. We emphasize that the choice of
measurement scheme does not affect the evolution of the system on
average. That is, for all measurement schemes, averaging over the possible
results and the possible values of the Rabi frequency yields the same
equation of motion for the system state. Physically, this is because
the average behaviour of the system is determined by its immediate
environment, whereas the different measurement schemes are effected
by  detecting the light emitted by the system in different ways.
However, the different measurement schemes give very different typical
posterior distributions, and very different rates of information gain.

Fourth, and perhaps most distinctively, we consider not just the
estimation of the unknown parameter, but also the estimation of the
state of the system conditioned on the measurement results \cite{perscom}.
We do this using the same Bayesian method as for the parameter
estimation. In this respect, our work could be
seen as an extension of quantum trajectory theory
\cite{Car93b} to systems with unknown dynamical parameters. Quantum
trajectory theory is simply the application of quantum measurement
theory to continuous monitoring of open quantum systems, most
usually optical systems subject to photodetection \cite{Wis96a}.

If the dynamical parameters for an open quantum system
are known then conditioning
the system on efficient detection of its emissions is guaranteed to
monotonically
increase its average purity in time, as information is gained about the
system. But if dynamical parameters are not known then the
average purity
may decrease, as the different possible evolutions are summed
incoherently. On the other hand, the measurement record also contains
information about these parameters, so that these parameters become
better defined over time. Hence one might expect that the
system will eventually become pure anyway.

It is one of the main results of this paper that this expectation is
not met. For our system there are
some monitoring schemes for which the parameter  never becomes
sufficiently well known for the system state to become pure. However,
there is no simple correlation between the information gained about
the parameter (the Rabi frequency) and the final purity of the system
(the atom). One monitoring scheme
yields almost no parameter information, yet produces, on average, a much
purer final system state than do other schemes that yield large
amounts of
parameter information. Moreover, the rates at which the system state
purifies is, for some monitoring schemes, tied to the rate of
parameter information gain, while for other monitoring schemes it is
much faster than that. These results can be understood only from an
appreciation of the conditional dynamics induced by the different
detection schemes.

The remainder of this paper is organized as follows. In Sec.~II we
present the general formalism for state and dynamical parameter
estimation by monitoring a single system. We also explain how the
parameter information gained is quantified. In Sec.~III we introduce
the system to which we apply our formalism,
a two-level atom, driven by an unknown Rabi frequency, and monitored
by having its fluorescence detected. Sec.~IV contains the results of
our numerical simulations of the relevant ensemble averages for five
different detection schemes: direct, the adaptive scheme of Wiseman
and Toombes \cite{WisToo99}, homodyne of the $x$
quadrature, homodyne of the $y$ quadrature, and heterodyne. Sec.~V
concludes.

\section{General Formalism}

\subsection{Quantum trajectories}

It is well known that quantum trajectories can be used to describe
the evolution of a continuously monitored open system \cite{Wis96a}.
Since here we are continuously monitoring an open
system with an unknown
dynamical parameter, we begin by giving a brief
outline of the standard quantum trajectory theory.

A good place to start is with the measurement formalism for open
systems \cite{Kraus83,BraKha92}. An open system is simply a
quantum system that interacts with its environment (usually called
a bath). This interaction, like all quantum interactions,
generally entangles the system and the bath. If we initially have
states $\ket{\psi(t_{0})}$ and $\ket{m(t_{0})}$ for the system and
bath respectively, and let these states entangle by $U(t_{0}+T)$,
a unitary operator which includes both the bath-system coupling
and the system dynamics. An instantaneous rank-one projective
measurement on the bath will result in the state after the
measurement being \beq
\ket{r}\ket{\psi_{r}(t_{0}+T)}=\frac{\ket{r} \bra{r} U(t_{0}+T
)\ket{m(t_{0})}\ket{\psi(t_{0})}}{\rt{{\rm P}(r)}}, \label{meas}
\eeq where P$(r)$ is the probability of getting the result $r$.
\erf{meas} shows that after the measurement the system and the
bath are disentangled, so it is not necessary to continue to
describe the bath in our treatment of the measurement. This allows
\erf{meas} to be reduced to \beq
\ket{\psi_{r}(t_{0}+T)}=\frac{{M}_{r}(T)\ket{\psi(t)}}{\rt{{\rm
P}(r)}}, \label{meas2} \eeq where
${M}_{r}(T)=\bra{r}U(t_{0}+T)\ket{m(t_{0})}$ is called the
measurement operator and has the feature of collapsing the
observer's knowledge of the system into a state that is consistent
with the result $r$. ${M}_{r}(T)$ is still an operator for the
system as $U(t_{0}+T)$ is an operator on the tensor product
Hilbert space for system and the bath. It is important to note
that this measurement operator is not necessarily a projector in
the system Hilbert space.

The probability P$(r)$ is given by
\beq
{\rm P}(r)={\rm Tr}[F_{r}(T)\ket{\psi(t_{0})}\bra{\psi(t_{0})}],
\label{prob}
\eeq
where $F_{r}$ is called the effect and is defined as
\beq
F_{r}(T)={M}_{r}\dg (T) { M}_{r}(T).
\label{POM}
\eeq
The complete set of effects must sum to one:
\beq
\sum_{r} F_{r}(t)=1.
\label{comp}
\eeq

The above formalism for measurement only considers pure states,
but to take into account initially mixed states
\erf{meas2} can be rewritten in terms of the state matrix.
The state after the measurement is then
\beq
\rho_{r}(t_{0}+T)={{M}_{r}(T)\rho(t_{0}) {M}_{r}
\dg(T)}/{{\rm P}(r)}.
\label{meas3}
\eeq
Here \erf{meas3} describes the state conditioned on the result $r$ and is
referred to as an {\it unraveling} of the average post-measurement
state
$\rho(t_{0}+T)$. That is, the weighted mean of all the possible conditioned
states for
one unraveling is equal to the average state:
\beq
 \rho(t_{0}+T)={\rm E}[\rho_{r}(t_{0}+T)]=\sum_{r}{\rm
 P}(r)\rho_{r}(t_{0}+T).
 \label{mean}
 \eeq
It should be noted that an average state has more
then one unraveling. The different unravelings correspond to
different sets of measurement operators, arising from different sets
of environment projectors $\ket{r} \bra{r}$ in \erf{meas}.

As mentioned earlier, quantum trajectories arises when this measurement
formalism is applied to a continuously monitored open system \cite{Wis96a}. In
continuous monitoring, repeated measurements of duration $\delta t$
are performed on the state. This results in the state being conditioned
on a record $\bfi$, which is a string containing the results $r_{k}$
of each  measurement. Here the subscript $k$ refers to
a measurement at time $t_{k}=k\delta t$, with $t_{0}=0$.
Using this $\bfi$, the conditioned state at time $t$
can be written as
\beq
\rhoc(t)={\tilde\rhoc(t)}/{{\rm P}(\bfi)},
\label{SME}
\eeq
where $\tilde\rhoc(t)$ is an unnormalized state conditioned on $\bfi$ and
is equal to
\bqa
&&\tilde\rhoc(t)\nl ={M}_{r_{k}}{M}_{r_{k-1}}\ldots
 {M}_{r_{1}}\rho(0) {M}_{r_{1}}
\dg\ldots {M}_{r_{k-1}}
\dg {M}_{r_{k}}
\dg.
\label{unp}
\eqa
The probability of obtaining this record is
\beq
{\rm P}(\bfi)={\rm P}(r_{k}){\rm
P}(r_{k-1})\ldots{\rm P}(r_{1})
={\rm Tr}[\tilde\rhoc(t)].
\label{recprob}
\eeq

To completely achieve continuous monitoring we let the time step
between measurements, $\delta t$, tend towards the infinitesimal
interval $dt$. In doing this, \erf{SME} defines a stochastic
master equation (SME), with its ensemble average reproducing the
usual deterministic master equation. That is, \bqa \rho(t)&&=
\sum_{\bfi} {\rm P}(\bfi) \rhoc(t)= \sum_{\bfi} \tilde\rhoc(t) \nl
=\lim_{\delta t \rightarrow 0} \sum_{r_{t/\delta t}\ldots r_{1}}
M_{r_{t/\delta t}} \ldots M_{r_{1}}\rho(0) M_{r_{1}}\dg \ldots
M_{r_{t/\delta t}}\dg  \nl =\lim_{\delta t \rightarrow 0}
\ro{1+{\cal L} \delta t}^{t/\delta t} \rho(0)=\exp({\cal L}
t)\rho(0), \eqa where for arbitrary $\rho$, ${\cal L}$ is the
Liouvillian superoperator defined as ${\cal L}\rho = \lim_{\delta
t \rightarrow 0} (\sum_{r}M_{r}\rho M_{r}\dg -\rho)/\delta t$.

\subsection{Quantum trajectories with an unknown parameter}

We now consider the situation where there is an unknown dynamical parameter
$\lambda$ in ${\cal L}$, and hence in the measurement operators $M_{r}$.
This is done by simply noting that for each $\lambda$
there will be a conditioned state. This gives a doubly conditioned state
of the form
\beq
\rho_{{\bf I},\lambda}(t)={\tilde\rho_{{\bf I},\lambda}(t)}
/{{\rm P}(\bfi|\lambda)},
\label{meas4}
\eeq
where ${\rm P}(\bfi|\lambda)$ is the probability of getting $\bfi$
{\it given} $\lambda$. It
is obtained by
\beq
{\rm P}(\bfi|\lambda)={\rm Tr}[\tilde\rho_{{\bf I},\lambda}(t)].
\label{recprob2}
\eeq

We wish to  determine the
posterior probability distribution ${\rm P}(\lambda|\bfi)$
of $\lambda$, given $\bfi$. This can be achieved using
 a Bayesian inference formula
\cite{Box73}.
\beq
{\rm P}(\lambda|\bfi)=\frac{{\rm P}(\bfi|\lambda){\rm
P}_{0}(\lambda)}{\int {\rm P}(\bfi|\lambda){\rm
P}_{0}(\lambda)d\lambda},
\label{bayes}
\eeq
where ${\rm P}_{0}(\lambda)$ is the prior distribution for
$\lambda$. For a ``good measurement'' of $\lambda$, as time increases, we
would
expect this prior distribution to converge to a $\delta$-distribution.

Theoretically, \erf{bayes} is complete for determining ${\rm
P}(\lambda|\bfi)$. However, in general ${\rm P}(\bfi|\lambda)$ is
very small and in numerical simulations it will incur large computer
roundoff errors. The small magnitude of ${\rm P}(\bfi|\lambda)$ is due to
the many possible trajectories the system could follow.

To overcome this problem, linear quantum trajectories \cite{GoeGra94b}
were used.
Linear quantum trajectories arise if we assume an ostensible
distribution for the result $r$, ${\rm \Lambda}(r)$ \cite{Wis96a}. These
${\rm \Lambda}(r)$ are independent of $\lambda$ and the only condition
 they must satisfy is that they add to one.
With these ostensible probabilities, the linear stochastic master equation
(LSME) is derived from \cite{Wis96a}
\beq
\bar{\rho}_{{\bf I},\lambda}(t)={\tilde\rho_{{\bf I},\lambda}(t)}
/{{\rm \Lambda}(\bfi)},
\label{plin}
\eeq
where the ostensible probability for getting $\bfi$
is
\beq
{\rm \Lambda}(\bfi)={\rm
\Lambda}(r_{k})
{\rm \Lambda}(r_{k-1})\ldots{\rm \Lambda}(r_{1}).
\eeq

The actual probability of getting $\bfi$ is \cite{Wis96a} \beq
{\rm P}(\bfi|\lambda)={\rm \Lambda}(\bfi){\rm Tr}[\bar{\rho}_{{\bf
I},\lambda}(t)]. \label{recprob3} \eeq Substituting ${\rm
P}(\bfi|\lambda)$ into \erf{bayes} we obtain \beq {\rm
P}(\lambda|\bfi)=\frac{{\rm Tr}[\bar{\rho}_{{\bf I},\lambda}
(t)]{\rm P}_{0}(\lambda)}{\int {\rm Tr}[\bar{\rho}_{{\bf
I},\lambda}(t)]{\rm P}_{0} (\lambda)d\lambda}. \label{bayes2} \eeq
From \erf{bayes2} we see that to calculate P$(\bfi|\lambda)$, the
norm of the linear conditioned state [\erf{plin}] is needed. The
order of magnitude of this norm is dependent on the ostensible
probability we chose. By \erf{recprob3}, if ${\rm \Lambda}(\bfi)$
is chosen to be of the same order as the true probability, this
norm will be of order unity. This avoids the problem of large
computer roundoff error.

\subsection{Quantifying the information gained}

One of the main aims of this paper is to classify the information gained
about the unknown parameter. The posterior probability calculated by
\erf{bayes2} contains all the information about $\lambda$ for a particular
record.
However the question remains, how can this information be quantified?
Two measures were investigated. The first is the variance:
\beq
{V_{{\bf I}}}=\int {\rm P}(\lambda|\bfi)\lambda^{2}
d\lambda
-\ro{\int {\rm P}(\lambda|\bfi)\lambda d\lambda }^{2}.
\label{var}
\eeq
The second is the information gain, $\Delta I_{{\bf I}}$
defined as \cite{Hall97}
\bqa
\Delta I_{{\bf I}}&=& \int {\rm P}(\lambda|\bfi) \log_{2}{\rm
P}(\lambda|\bfi)d\lambda \nl
 -\int {\rm P}_{0}(\lambda) \log_{2}{\rm
P}_{0}(\lambda)d\lambda.
\label{bits}
\eqa
This measures the number of bits of
information gained by the  observer about the
parameter $\lambda$. It can be thought of as the negative change in
entropy of $\lambda$.
The greatest information gain
corresponds to the transition from a flat (most disordered) distribution
to a peaked (most ordered) distribution.

These parameters give an indication of the quality of knowledge
gained by an observer, for a particular run of the experiment. To
characterize a particular measurement scheme, it is necessary to
calculate the ensemble averages of $V_{{\bf I}}$ and $\Delta
I_{{\bf I}}$, which we denote as $V$ and $\Delta I$. The ensemble
average of a parameter $A_{{\bf I}}$ is defined as \bqa A&&={\rm
E}[A_{{\bf I}}]=\sum_{\bfi} A_{{\bf I}}{\rm P}(\bfi) \nl
=\sum_{\bfi}\int A_{{\bf I}}{\rm P}(\bfi|\lambda) {\rm
P}_{0}(\lambda)d \lambda. \eqa

Numerically, this is done by picking a true $\lambda$,
$\lambda_{\rm true}$, randomly from P$_{0}(\lambda)$, and then
simulating a quantum trajectory for this $\lambda_{\rm true}$,
yielding $\bfi$. This gives a typical record as would be obtained
experimentally. This $\bfi$ is then used to calculate
Tr$[\bar\rho_{{\bf I},\lambda}(t)]$ for all $\lambda$'s in the
range of P$_{0}$. This allows the calculation of
P$(\lambda|\bfi)$, with this probability the parameter of interest
$A_{{\bf I}}$ can be calculated. By storing this value and
repeating the above procedure $n \gg 1$ times,  the ensemble
average $A$ of
 $A_{{\bf I}}$ is obtained.

\subsection{Best estimate of conditioned state}

Another aim of this paper was to determine the best estimate
 of the state given the
knowledge we have obtained from a measurement. In \erf{meas4} we defined the
doubly conditioned state that arose when the state was conditioned on both
$\bfi$ and $\lambda$. From \erf{meas4} there are two best estimate
states that can be calculated. They are $\rho_{\lambda}$ and
$\rhoc$ and can be interpreted as the best
estimate state, when $\lambda$ or $\bfi$ is known respectively. They are
defined as follows
\bqa
\rho_{\lambda}(t) &=&\sum_{\bfi}\tilde\rho_{{\bf I},\lambda}(t),
\label{best1} \\
\rho_{{\bf I}}(t)&=&\frac{\int\tilde\rho_{{\bf I},\lambda}
(t){\rm P}_{0}(\lambda)d\lambda }
{\int {\rm P}(\bfi|\lambda) {\rm P}_{0}(\lambda)d\lambda}.
\label{best2}
\eqa
It should be noted that the average of each of these states will give the
same average state $\rho(t)$.

\Erf{best1} describes the best estimate state that arises when the
dynamical parameter is known and the record is not (i.e. a non-monitored
system). This obeys master equation $\dot
\rho_{\lambda}={\cal L}_{\lambda}\rho_{\lambda}$. Of more interest to us is the
best estimate state described by \erf{best2}, which is
the state conditioned on some observed record $\bfi$,
when the true value of $\lambda$ is unknown.

In calculating $\rho_{{\bf I}}$, if we use \erf{best2}, we again run into
the problem that
 the magnitude of $\tilde\rho_{{\bf I},\lambda}(t+dt)$ will typically be
very small. Again this is overcome by using linear quantum
trajectories, replacing  \erf{best2} by
\beq
\rho_{{\bf I}}(t)=
\frac{\int\bar\rho_{{\bf I},\lambda}
(t){\rm P}_{0}(\lambda)d\lambda }
{\int {\rm Tr}[\bar\rho_{{\bf I},\lambda}
(t)] {\rm P}_{0}(\lambda)d\lambda}.
\label{best3}
\eeq

To quantify the information gained about the state, the purity ($p_{{\bf
I}}$) can be determined,
\beq
p_{{\bf I}}={\rm Tr}[\rho_{{\bf I}}(t)^{2}].
\eeq
The
ensemble average purity
($p={\rm E}[p_{{\bf I}}]$) will give us an indication of how
well the measurement scheme is at producing pure states. One might expect
that a high $p$ would correspond to a high $\Delta I$.
However it will be seen that this is not true.

\section{The System}

The system we are considering is a classically driven two level atom,
 immersed in the vacuum. With no monitoring of the
vacuum field, the average state evolution when all the dynamical
parameters are known is given by the master equation. The Lindblad
form \cite{Lin76} of the master equation for the TLA, in the
interaction picture (with respect to the free evolution of the
atom) is \cite{WalMil94} \beq
\dot\rho(t)=-\frac{i\Omega}{2}[\sigma_{x},\rho(t)] +\gamma{\cal
D}[\sigma]\rho(t)={\cal L}_{\Omega}\rho(t). \label{master} \eeq
Here $\Omega$ is the Rabi frequency, $\gamma$ is the spontaneous
emission rate, $\sigma$ is the lowering operator, $\sigma_{x}$ is
the usual Pauli matrix and ${\cal D}$ is the superoperator that
represents damping of the system into the environment. It is
defined as \cite{WisMil93c} \beq {\cal D}[a]\rho=a\rho a\dg
-\half\cu{a\dg a\rho+ \rho a\dg a}. \eeq The solution of this
equation can be described by the Bloch vectors $(x,y,z)$, with
$\rho$ written as \beq
\rho=\half(1+x\sigma_{x}+y\sigma_{y}+z\sigma_{z}). \eeq The purity
$p$ is equal to \beq p=\half(1+x^{2}+y^{2}+z^{2}), \eeq

Using this Bloch representation the solution of \erf{master} is a state
that rotates about the $x$-axis at frequency $\Omega$, with damping in
all variables towards the steady
state value of
\beq
x_{\rm ss}=0,\hspace{0.2in}
y_{\rm ss}=\frac{2\Omega\gamma}{2\Omega^{2}+\gamma^{2}},\hspace{0.2in}
z_{\rm ss}=\frac{-\gamma^{2}}{2\Omega^{2}+\gamma^{2}}.
\label{steady}
\eeq

The most obvious choice for the unknown dynamical parameter is
$\Omega$, as indicated by the subscript in ${\cal L}_{\Omega}$ in
\erf{master}. This can be physically motivated as follows: if we
placed a laser-cooled atom (with no center-of-mass motion)
 in a classical standing field, then the $\Omega$ it
would experience is
\beq
\Omega=\Omega_{\rm max}\sin(kx),
\label{omega}
\eeq
where $k$ is the wavevector for the classical field and $x$ is the
position of center of mass of the atom. We assume that the
placement of the atom in the field is not biased in any way. That is, in
one wavelength ($\lambda$) of the field the atom position
distribution is given by ${\rm P}_{0}(x)=1/\lambda$.
Using \erf{omega}, ${\rm P}_{0}(x)$ can be transformed into a
probability distribution
in $\Omega$ space,
\beq
{\rm P}_{0}(\Omega)=\frac{1}{\pi\sqrt{\Omega_{\rm max}^{2}-\Omega^{2}}}.
\label{omegadis}
\eeq
This is the prior distribution for $\Omega$, that will be used in the
 rest
of this paper, with $\Omega_{\rm max}=10\gamma$. Along with this prior
distribution the initial condition that we will use for our
simulations,
unless otherwise stated, is $\rho_{{\bf I},\Omega}(0)$ satisfying
\beq
{\cal L}_{\Omega} \rho_{{\bf I},\Omega}(0)=0.
\eeq
That is, we will assume the initial state is the steady state of the
general master equation \erf{master}.

\section{Results}

The results of this paper are broken down into five subsections,
each corresponding to one of the five measurement
schemes investigated.

\subsection{Direct Detection} \label{direct}

The first measurement scheme investigated was direct detection.
This
 involves the detection of all the fluorescence emitted
by the atom as shown in Fig.~1. Continuous monitoring with this
detection scheme will yield either one of two results for each
interval $dt$, a detection (labelled by a 1) or no detection
(labelled by a 0). Thus $\bfi$ will be a string of 0's and 1's.
The measurement operators for each of these results are
\cite{Wis96a} \bqa
{M}_{1}(dt)&=&\rt{dt\gamma}\sigma, \\
{M}_{0}(dt)&=&1-\ro{i\frac{\Omega}{2}\sigma_{x}+
\frac{\gamma}{2}\sigma\dg\sigma}dt.
\eqa
It can be shown that these measurement operators satisfy the
completeness condition, \erf{comp}. Using these
measurement operators and
\erf{SME}, a SME for direct detection can written as
\beq
d\rho_{{\bf I},\Omega}=dN(t){\cal G}[\rt{dt\gamma}\sigma]\rho_{{\bf
I},\Omega}-
dt{\cal H}[i\frac{\Omega}{2}\sigma_{x}+
\frac{\gamma}{2}\sigma\dg\sigma]\rho_{{\bf I},\Omega},
\label{stodi}
\eeq
where ${\cal G}$ and ${\cal H}$ are the nonlinear superoperators
defined for arbitrary $a$ and $\rho$ by
\bqa
 {\cal G}[a]\rho&=&\frac{a\rho a\dg}{{\rm Tr}[a\rho a\dg]}-\rho, \\
 {\cal H}[a]\rho&=&a\rho+\rho a\dg -{\rm Tr}[a\rho +\rho a\dg]\rho.
\eqa

\begin{figure}[t]
\includegraphics[width=0.45\textwidth]{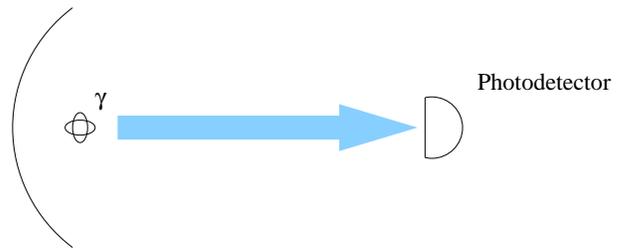}
\caption{\label{fig1} A schematic for direct detection. The atom
is placed at the focus of a parabolic mirror so that all the
fluorescence emitted by the atom is detected by the
photodetector.}
\end{figure}

In \erf{stodi}, the variable $dN$ is a stochastic increment that
equals one if there is a
detection in the interval $dt$ and equals 0 otherwise. Formally, $dN$
is defined by
\bqa
 dN(t)^{2}&=&dN(t), \\
 E[dN(t)]&=&{\rm P}(1)=dt \gamma \an{\sigma\dg\sigma}.
\label{epect}
\eqa

By averaging \erf{stodi} and using \erf{epect},
it is easily seen that the SME is an unraveling of the general master
equation, \erf{master}. A typical trajectory of this SME is shown in
Fig.~2 (solid line), for $\Omega=5\gamma$. It is observed that the $x$
component is zero, and the $y$ and $z$ oscillate in
quadrature. This can be understood
physically as the state is dominated by the $\Omega \sigma_{x} /2$
Hamiltonian, with detections occurring stochastically according to
\erf{epect}. After
each detection the  state collapses to the ground
state ($x=0$, $y=0$, and $z=-1$).

\begin{figure}[t]
\includegraphics[width=0.45\textwidth]{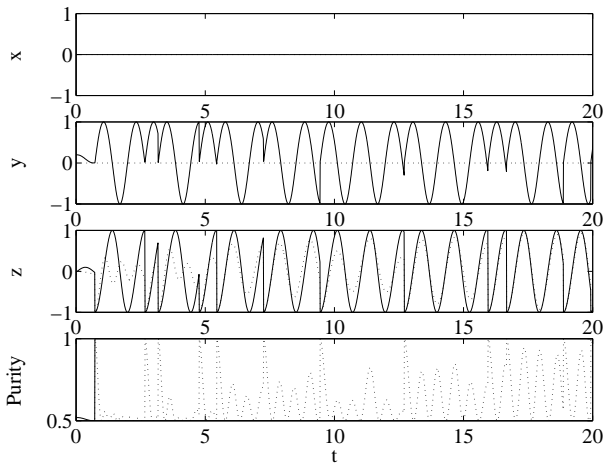}
\caption{\label{fig2} The best estimate states when $\Omega$ is
known (Solid line) and unknown (dotted line) for $\Omega_{{\rm
true}}=5\gamma$ when direct detection is used. Time is measured in
units of $\gamma^{-1}$, $x$, $y$, and $z$ are the Bloch vector
components and the purity $p=\half(1+x^{2}+y^{2}+z^{2})$. The
initial states are the steady state for the known case and the
average steady state for the unknown case.}
\end{figure}

To consider the case when $\Omega$ is unknown, a LSME had to be developed.
Using the direct detection measurement operators and \erf{plin}, with
$\Lambda(r)$ defined as
\beq
\Lambda(1)= \epsilon dt = 1-\Lambda(0),
\eeq
where $\epsilon$ is an arbitrary parameter, the LSME is
\bqa
d\bar\rho_{{\bf I},\Omega}&=&dN(t)\bar{\cal
G}[\rt{dt\gamma}\sigma]\bar\rho_{{\bf I},\Omega} \nl -
dt\bar{\cal H}[i\frac{\Omega}{2}\sigma_{x}+
\frac{\gamma}{2}\sigma\dg\sigma-\frac{\epsilon}{2} ]\bar\rho_{{\bf
I},\Omega}.
\label{lstodi}
\eqa
The $\bar{\cal G}$ and $\bar{\cal H}$ linear superoperators are
defined as
\bqa
\bar{\cal G}[a]\bar\rho &=&\frac{a\bar\rho a\dg}{\epsilon dt}-\bar\rho, \\
\bar{\cal H}[a]\bar\rho &=&a\bar\rho+\bar\rho a\dg.
\eqa

To obtain the general master equation from \erf{lstodi},
E$[dN]=\Lambda(1)$ has to be used.  However, to determine the
parameters of interest to us, namely $\rho_{\bf I}(t)$ and
P$(\Omega|\bfi)$, \erf{lstodi} is numerically simulated for all
possible $\Omega$ in P$_{0}(\Omega)$ with $dN$ specified by
$\bfi$.  This record would ideally be obtained experimentally but
for the purpose of this paper it is calculated by numerically
evaluating \erf{stodi} for a known $\Omega$, which we will refer
to as $\Omega_{\rm true}$.  This $\bfi$ is then used in
\erf{lstodi} to generate ${\rm Tr}[\bar\rho_{{\bf I},\Omega}]$ for
all the $\Omega$'s between $-\Omega_{\rm max}$ and $\Omega_{\rm
max}$. Then with \erf{best3} and \erf{bayes2} one can obtain both
$\rho_{\bf I}(t)$ and P$(\Omega|\bfi)$.

For a $\bfi$ based on $\Omega_{\rm true}=5\gamma$
the best estimate state and the posterior distribution where
calculated and are shown in Fig.~ 2 and 3 respectively.  It is observed
that, in contrast to the known $\Omega$ case, the best estimate of
$y$ is identically zero. This is because positive and negative
$\Omega$ are initially equally likely, so that $y_{\rm ss}$ in
\erf{steady}
averages to zero. Moreover, the sign of $\Omega$ is not determinable
by this measurement scheme, because the rate of detections depends
only on $z$, which is independent of the sign of $\Omega$.

\begin{figure}[t]
\includegraphics[width=0.45\textwidth]{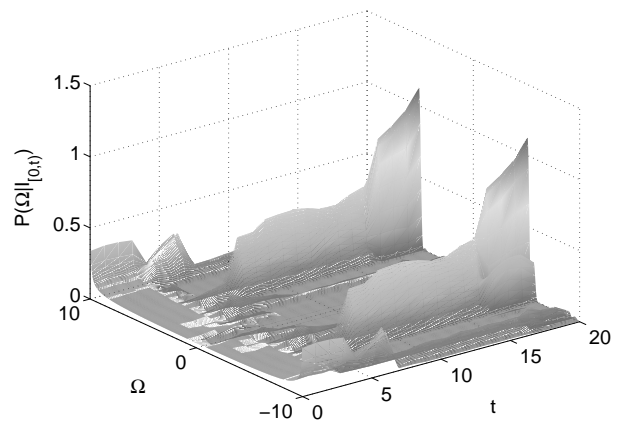}
\caption{\label{fig3} A plot of a typical ${\rm P(\Omega|\bfi)}$
when $\Omega_{{\rm true}}=5$ for the direct detection scheme.
$\Omega$ is measured in units of $\gamma$ and time is measured in
units of $\gamma^{-1}$.}
\end{figure}

Another difference apparent with the unknown $\Omega$ case is that
$z$ oscillates with a different frequency to the known $\Omega$ case,
in this case a faster frequency [since $P_{0}(\Omega)$ is peaked at
the end points $|\Omega| = \Omega_{\rm max} = 10 \gamma$].
However as time increases its frequency tends to that of
the known case.  This is
due to the fact that for direct detection the rate of detections is
dependent on the magnitude of $\Omega$, so as time goes on
one would expect to gain more information about the magnitude of
$\Omega$.

These interpretations of the conditioned dynamics are confirmed in
Fig.~3.  With increasing time, the posterior distribution
localizes at $\pm\Omega_{\rm true}$. The mean is always zero and
thus is not an unbiased estimator of $\Omega$. The reason that the
magnitude is determinable and the sign is not, can be formulated
as follows. In the Bloch representation of \erf{lstodi}, with the
transformation $y\rightarrow -y$, $\Omega \rightarrow -\Omega$ the
equations stay invariant. Since this transformation changes the
direction of rotation around the $x$-axis, we will call it
 the rotation transformation.

With an indeterminable
direction of rotation and this measurement scheme,
it can be seen that the best estimate state will never become
more pure than a state that is a mixture of two states that rotate is
opposite directions around the $x=0$ great circle of the  Bloch sphere.
Thus the best estimate state oscillates up and down the
$z$-axis of the Bloch sphere.

We turn now to quantifying the measurement scheme's ability to gain
knowledge, by  numerically determining
 the ensemble average purity, $V$ and $\Delta
I$.  These ensemble averages were calculated for $\Omega_{\rm
true}$'s weighted on the prior distribution, \erf{omegadis}.
These numerical simulations are depicted in Fig.~4 for two initial
states; one is the steady state (solid line) and the other is the
ground state (dotted line).  It is observed that in both cases the
average purity of the state never attains one, with the purity in
the second case initially decreasing from one.  The long time
purity ($\simeq 0.75$) is due the best estimate being a mixture of
two states as explained above.  This figure can be obtained
analytically, if we make the follow two assumptions. The first is
that $\Omega_{\rm true}\gg \gamma$. This is valid as
$P_{0}(\Omega)$ from which $\Omega_{\rm true}$ is drawn  is peaked
at $\pm \Omega_{\rm max}$, and in our calculations $\Omega_{\rm
max} \gg\gamma$. The second assumption is that in the long time
limit the  posterior distribution localizes on $\pm\Omega_{\rm
true}$, which is what is seen in Fig.~3.  With these two
assumption the long-time best estimate state in Bloch
representation will be \beq x=0, \hspace{0.2in} y=0,
\hspace{0.2in} z\simeq-\cos\Omega_{\rm true}(t-t_{\rm last}), \eeq
where $t_{\rm last}$ is the time of the last jump, which is
typically more than one Rabi cycle before $t$. With this state the
average purity (for the long time limit) can be estimated as \beq
p\space=
\frac{\Omega}{2\pi}\int_{0}^{2\pi/\Omega}\frac{1+z(s)^{2}}{2} ds
=\frac{3}{4}. \eeq

\begin{figure}[t]
\includegraphics[width=0.45\textwidth]{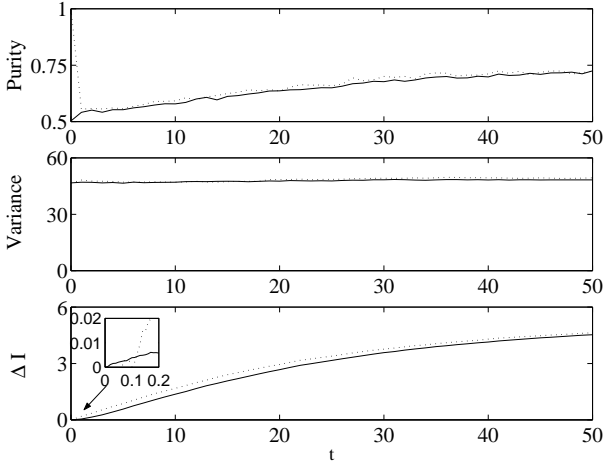}
\caption{\label{fig4} The ensemble average ($n=1000$) of the
purity, variance and $\Delta I$ when direct detection is used, for
two initial states, the steady state (solid) and ground state
(dotted). Time is measured in units of $\gamma^{-1}$.}
\end{figure}

From Fig.~4, it is also observed that the simulated
ensemble average variance $V$ is approximately
constant for all time. In fact, given that the no information about
the sign of $\Omega$ is
 determinable, and that the initial distribution ${\rm P}_{0}(\Omega)$ is
 symmetric, it is easy to prove that $V$ is exactly constant.

For the third parameter $\Delta I$, it
is observed that, on average, direct detection yields information
about $\Omega$ as time increases, for both initial states. It is
observed that the initial slope of $\Delta I$ is zero for the ground
state, while it is non-zero
for the initial steady state case. The initial flatness in the first
case is due to the fact that if the system starts in the ground, the
rate of detections (proportional to the excited state component)
scales as $(\Omega_{\rm true}t)^{2}$,
 and with out any detections it would not be possible
to gain any information. By contrast, for the steady state case there will be
some excited state fraction (depending on $\Omega$)
and thus a finite detection rate even at
$t = 0$. Fig.~4 also show that, after the initial flatness, the  $\Delta
I$ in the first case rapidly overtakes that in the second case.
This jump in $\Delta
I$ occurs at roughly $t=1/\Omega_{\rm max}$, which is when one would expect
a significant excited state fraction to have developed (Recall that
${\rm P}_{0}(\Omega)$ is sharply peaked at $\Omega = \pm \Omega_{\rm max}$).

\subsection{Adaptive Detection} \label{adapt}

The second measurement scheme investigated was the
adaptive scheme of Wiseman and Toombes \cite{WisToo99}. For a known
$\Omega$, this
measurement scheme is designed to keep the atom jumping between two fixed
states. For $\Omega$ large, these fixed states turn out to be close to
$\sigma_{x}$ eigenstates. This two-state jumping
is achieved by coherently mixing
the fluorescence emitted
from the atom with a weak local oscillator (LO) via a low-reflectance
 beam splitter (see Fig.~5).
The reflected
amplitude $\mu$  of the local oscillator is switched between
$\pm\half\sqrt{\gamma}$ each time a
detection is registered by the photodetector.

\begin{figure}[t]
\includegraphics[width=0.45\textwidth]{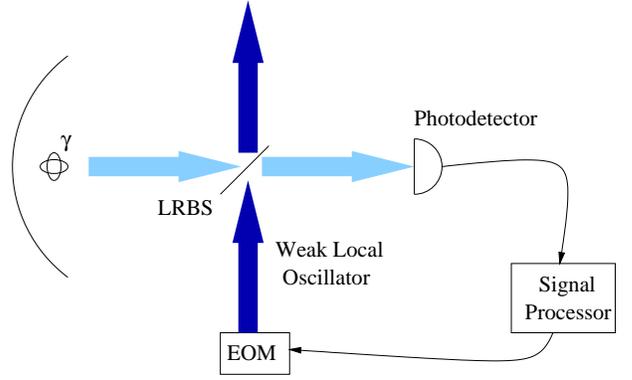}
\caption{\label{fig5} A schematic for adaptive detection. The
fluorescence emitted by the atom is coherently mixed with a weak
local oscillator (LO) via a low reflectivity beam splitter (LRBS).
The electro-optic modulator (EOM) reverses the amplitude of the
 LO every time the photodetector fires.}
\end{figure}

For this detection scheme the measurement operators are \cite{WisToo99}
\bqa
M_{1}(dt)&=&\rt{dt\gamma}(\sigma+\mu), \\
{
M}_{0}(dt)&=&1-(i\frac{\Omega}{2}\sigma_{x}+
\frac{\gamma}{2}\sigma\dg\sigma+\mu\gamma\sigma+\frac{\gamma\mu^{2}}{2})dt.
\nl
\eqa
These measurement operators result in a SME of the form
\beq
d\rho_{{\bf I},\Omega}=dN(t){\cal
G}[\rt{dt\gamma}(\sigma+\mu)]\rho_{{\bf I},\Omega}
-dt{\cal H}[\zeta]\rho_{{\bf I},\Omega},
\label{stoad}
\eeq
 where
 \beq
 \zeta=
i\frac{\Omega}{2}\sigma_{x}+
\frac{\gamma}{2}\sigma\dg\sigma+\mu\gamma\sigma+\frac{\gamma\mu^{2}}{2}.
\eeq
Using the same ostensible distribution ${\rm \Lambda}(r)$ as in direct
detection, the
LSME is
\beq
d\bar\rho_{{\bf I},\Omega}=dN(t)\bar{\cal
G}[\rt{dt\gamma}(\sigma+\mu)]\bar\rho_{{\bf I},\Omega}-
dt\bar{\cal H}[\bar\zeta]\bar\rho_{{\bf I},\Omega},
\label{lstoad}
\eeq
where
\beq
\bar\zeta=i\frac{\Omega}{2}\sigma_{x}+
\frac{\gamma}{2}\sigma\dg\sigma+\mu\gamma\sigma+\frac{
\gamma\mu^{2}}{2}-\frac{\epsilon}{2}.
\eeq

Figure 6 shows the best estimate state for a known (solid) and
unknown $\Omega$ (dotted), with $\Omega_{\rm true}=5\gamma$. It is observed
that with the
known $\Omega$ case after the initial transients, the state jumps
between the  two fixed states \cite{WisToo99}
\beq
 x=\frac{\mp2\Omega^{2}}{2\Omega^{2}+{\gamma^{2}}},\hspace{0.2in}
y=\frac{2\Omega\gamma}{2\Omega^{2}+\gamma^{2}},\hspace{0.2in}
z=\frac{-\gamma^{2}}{2\Omega^{2}+\gamma^{2}}.
\label{adap}
\eeq
For the
unknown $\Omega$ case the $y$ component averages to zero, and the $x$ and $z$
components
both appear to be slightly different to the known $\Omega$ case.

\begin{figure}[t]
\includegraphics[width=0.45\textwidth]{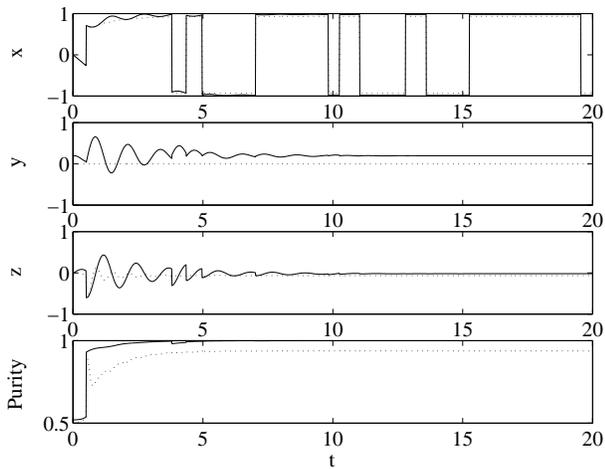}
\caption{\label{fig6} The best estimate states when adaptive
detection is used. Details are as in Fig.~2.}
\end{figure}

Similarly to the direct detection case, a better
understanding of this state can be obtained by considering ${\rm
P}(\Omega|\bfi)$. This is shown in Fig.~7 and it can be seen
that as time increases under this adaptive measurement, the typical
posterior probability distribution ${\rm P}(\Omega|\bfi)$ scarcely changes
 from ${\rm P}_{0}(\Omega)$.
This is not unexpected, as for this detection scheme it can be shown
that at steady state the jumps are Poissonian, with rate $\gamma/4$.
That is, the jumps are independent of
$\Omega$ \cite{WisToo99} and hence yield no information about it.
Since P$(\Omega|\bfi) \approx {\rm P}_{0}(\Omega)$, we can use this
approximation to obtain analytically
an indication of the best estimate state by solving
\erf{best2}.
For this detection scheme this is simply the mean of
\erf{adap} under the distribution ${\rm P}_{0}(\Omega)$. This gives
\beq
 x=\mp(1-\frac{
\gamma^{2}}{\sqrt{2\Omega_{\rm max}^{2}+\gamma^{2}}}),\hspace{0.1in}
y=0,\hspace{0.1in}
z=\frac{-\gamma}{\sqrt{2\Omega_{\rm max}^{2}+\gamma^{2}}}.
\label{adap2}
\eeq
Comparing this with the numerical simulation it is observed that they agree
very well.

\begin{figure}[t]
\includegraphics[width=0.45\textwidth]{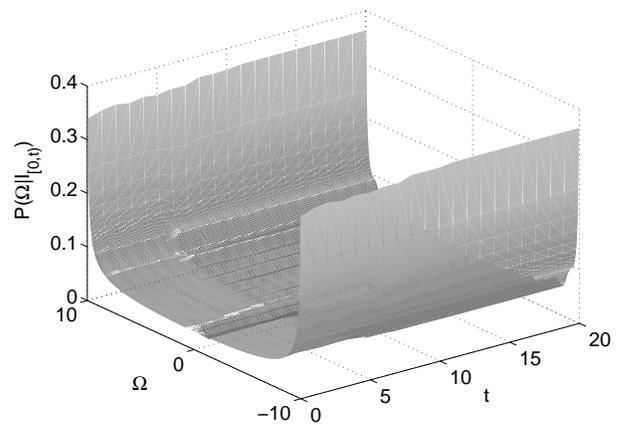}
\caption{\label{fig7} A plot of ${\rm P(\Omega|\bfi)}$
 for the adaptive scheme. Details are as in Fig.~3.}
\end{figure}

To quantify this detection scheme, the ensemble average of the variance,
purity and $\Delta I$ were numerically calculated and are
shown in Fig.~8. The purity rapidly becomes,
and remains,
relatively high. This is because the best estimate state of
\erf{adap2} is
the same no matter what $\Omega_{\rm true}$ is chosen. For
$\Omega_{\rm max}=10\gamma$ the numerical value of the stationary
purity is 0.934 and by
using \erf{adap2} an analytical value of the purity can be obtained,
\beq
p=1 + \frac{{\gamma }^2}{{\gamma }^2 + 2\,{\Omega_{\rm max} }^2} -
  \frac{\gamma }{{\sqrt{{\gamma }^2 + 2\,{\Omega_{\rm max} }^2}}}.
\eeq
For $\Omega_{\rm max}=10\gamma$ this gives a value of 0.934, which is
equal to the numerical value.

\begin{figure}[t]
\includegraphics[width=0.45\textwidth]{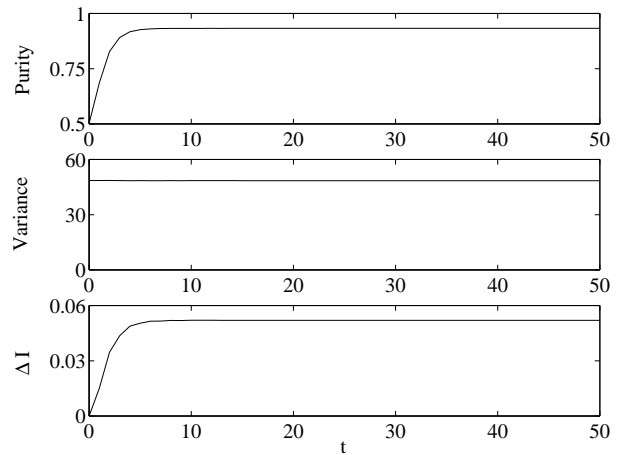}
 \caption{\label{fig8}The ensemble average
($n=1000$) of the purity, variance and $\Delta I$ when the
adaptive detection technique was used. Note for $\Delta I$ the
scale has been change when compared to Fig.~4. }
\end{figure}

Since this state has a high purity one might expect that the
unknown parameter must also be well defined.  However this is not
true as already discussed.  This lack of knowledge about $\Omega$
is seen in Fig.~8. Like direct detection, the sign of $\Omega$
cannot be determined so the average variance remains precisely
constant. However unlike direct detection, the information gain is
bounded, with a maximum $\Delta I$ of less than 0.06 bits.

An interesting point to note is this scheme would be well suited to
estimating $\gamma$ (if there were some uncertainty in that parameter)
even if $\Omega$ was also uncertain. That is because the detection
rate is proportional to $\gamma$, almost independent of $\Omega$. Of course
this would require the local oscillator amplitude to be adjusted
from an initial guess according to the best estimate of $\gamma$.

\subsection{Homodyne $x$ Detection} \label{homox}

To perform a homodyne detection experiment, a similar arrangement to the
adaptive scheme is used. That is, the output flux from the
atom is mixed with a resonantly tuned LO by a beam splitter (see Fig.~9).
However in this scheme there is no feedback and the amplitude $\beta$ of
the LO
is assumed to be infinite ($\beta\rightarrow\infty$). Because of this,
there will be many detections in the interval $dt$. Each detection
causes only an infinitesimal change in the system state, so the
evolution of the system can be described by a diffusive SME. In
each $dt$ there will be a continuous current $I$ registered in $\bfi$
rather than a detection or no-detection. Since $I$ is a continuous
variable we can define a measurement operator, $M_{I}$, a continuous
function of $I$,
to represent this measurement scheme,
\beq
M_{I}=\sqrt{\Upsilon_{I}}[1-(i\frac{\Omega}{2}\sigma_{x}+
\frac{\gamma}{2}\sigma\dg\sigma-\rt{\gamma}\sigma
e^{-i\Phi}I)dt].
\label{homo}
\eeq
Here $\Phi$ is the phase of the local oscillator and
\beq
\Upsilon_{I} dI=\frac{1}{\rt{2\pi/dt}}e^
{-\half I^{2}dt} dI,
\eeq
is a Gaussian probability measure.
It is easily shown that this continuous measurement operator
satisfies the completeness
condition, \erf{comp}, where the sum is replaced by an integral over
$I$
between $\pm\infty$.

\begin{figure}[t]
\includegraphics[width=0.45\textwidth]{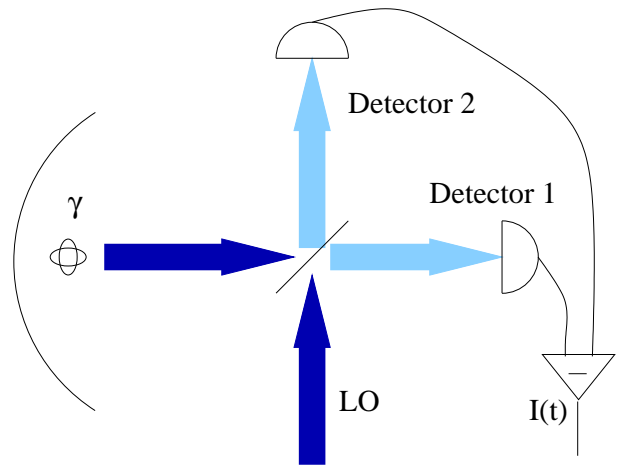}
 \caption{\label{fig9} A schematic for the three
detection schemes,  homodyne of the $x$ quadrature, homodyne of
the $y$ quadrature and heterodyne. For the homodyne schemes the LO
is resonantly tuned to the atomic frequency, with a phase of zero
and $\pi/2$
 for the $x$
 and $y$ schemes respectively, whereas for the heterodyne it is detuned
 by an amount
$\Delta$.}
\end{figure}

With this continuous measurement operator the SME in the \ito form is
\cite{SME}
\bqa
d\rho_{{\bf I},\Omega}
&=&{\cal L}_{\Omega}\rho_{{\bf I},\Omega}dt
 +\rt{\gamma}{\cal H}[\sigma
e^{-i\Phi}]\rho_{I}\nl \times(Idt-\rt\gamma{\rm Tr}[\sigma
e^{-i\Phi}\rho_{{\bf I},\Omega}+\rho_{{\bf I},\Omega}\sigma\dg
e^{i\Phi}]dt),\nl \label{stoho} \eqa where $I$ is the current
element for the interval $dt$ and is equal to the difference
between the number of detections at the two photodiodes divide by
the intensity of the field. By using \erf{homo} and \erf{prob} the
probability of getting $I$ for the interval  $dt$ can be
calculated. This gives a Gaussian distribution with a mean equal
to $\rt{\gamma}\an{\sigma e^{-i\Phi}+\sigma\dg e^{i\Phi}}$ and a
variance of $dt^{-1}$. Thus, $I$ will be a Gaussian random
variable (GRV) of the form \beq I= \rt{\gamma}{\rm Tr}[\sigma
e^{-i\Phi}\rho_{\bf I}+\rho_{\bf I}\sigma\dg e^{i\Phi}]+ \xi (t),
\eeq where $\xi (t)=dW(t)/dt$ represents Gaussian white noise, and
is formally defined as \cite{Gar85} \beq {\rm E}[\xi(t)] = 0
\;,\;\; {\rm E}[\xi(t)\xi(t')] = \delta (t-t'). \label{noise} \eeq

For the LSME we take the ostensible probability for the current to
be equal to that which would arise from the LO alone. This results
in ${\rm \Lambda}(I)= \Upsilon_{I}$ so that $I$ is ostensibly a
GRV with mean zero and variance $dt^{-1}$, like $\xi(t)$. The LSME
in \ito form is \cite{SME} \beq d\bar\rho_{{\bf I},\Omega} ={\cal
L}_{\Omega}\bar\rho_{{\bf I},\Omega}dt +\rt{\gamma}\bar{\cal
H}[\sigma e^{-i\Phi}]\bar\rho_{{\bf I},\Omega} Idt. \label{lstoho}
\eeq

It can be seen that both the LSME and the SME reduce to \erf{master}
when the ensemble average is taken. Similarly to the
 previous schemes, to determine an
unknown $\Omega$, $\bfi$ is generated by the
SME for a preset $\Omega$, $\Omega_{\rm true}$, which may  then be
``forgotten''. The LSME is then used to generate both
$\rho_{I}(t)$ and P$(\Omega|\bfi)$ for the predetermined record $\bfi$.

For homodyne $x$ quadrature measurement, the $\Phi$ of the LO is set
to zero (as $x=\an{\sigma+\sigma\dg}$).  With this phase and
$\Omega_{\rm true}=5\gamma$, the best estimate state for a known and
unknown $\Omega$ are shown in Fig.~10.  It is observed that for the
known $\Omega$ case, the state seems to localize itself relatively
fast into pure states that have a large $x$ contribution, and small
oscillations in the $y$ and $z$ directions.  By contrast, when
$\Omega$ is unknown, the best estimate state still contains a large
$x$ contribution, but the $y$ is strictly zero and the amplitude of the
$z$ oscillations is reduced.  As in the previous cases,
 this zero $y$ component can be
understood by considering ${\rm P}(\Omega|\bfi)$, shown in Fig.~11.
It is seen that, like direct detection, this measurement scheme has an
even posterior distribution that localizes at $\pm\Omega_{\rm true}$.
This is again due to the stochastic
Bloch equations being invariant under the
previously considered rotational transformation.  However, the rate at
which this localization occurs is much slower than under direct
detection.

\begin{figure}[t]
\includegraphics[width=0.45\textwidth]{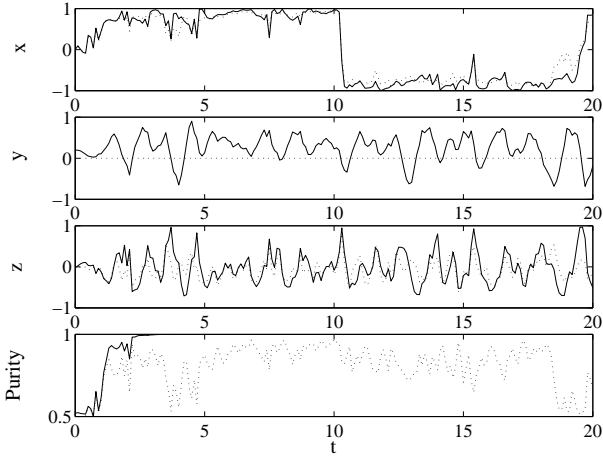}
\caption{\label{fig10} The best estimate states for  the homodyne
$x$ scheme. Details are as in Fig.~2.}
\end{figure}

\begin{figure}[t]
\includegraphics[width=0.45\textwidth]{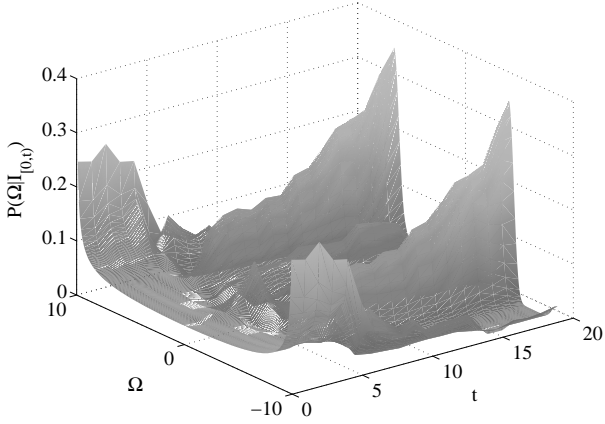}
\caption{\label{fig11} A plot of ${\rm P(\Omega|\bfi)}$ for the
homodyne $x$ scheme. Details are as in Fig.~3.}
\end{figure}

The slower rate of information gain is confirmed with the
calculation of the ensemble average of $\Delta I$, shown in Fig.~12. It is
seen that within $50\gamma^{-1}$ units of time, $\Delta I$ for homodyne $x$ is
about half that of direct detection. Physically this comes about because,
for the system we are investigating, the underlying dynamics cause
the states to rotate around the $x$-axis with frequency
$\Omega_{\rm true}$. The measurement scheme tends to produce states
oriented mainly in the $\pm x$ directions. This can be understood from
the measurement effect $F_{I}$, which, using \erf{POM}, can be shown to
be
 \beq
F_{I} dI= \frac{1}{\sqrt{2\pi/dt}} e^{-\half (I-\rt{\gamma}\sigma_{x})^{2}dt}dI.
\eeq
This effect is a Gaussian with a mean equal to the $\sigma_{x}$ quadrature
operator and variance $dt^{-1}$. Thus, it is an unsharp
measurement of $\sigma_{x}$. Thus, for a
measurement scheme that makes the conditioned state
mainly oriented in the $\pm x$ directions,
one would expect that this state would be
less affected by an unknown $\Omega$ than a state on the $x=0$ plane
as produced by direct detection. Thus less information about $\Omega$
comes out of the measurement record.
In Fig.~11 it is observed that the ensemble average of the purity of
this state increases
quickly to about 0.75, then increases only slowly afterwards. This quick
increase is also a result of the state becoming
predominately $\pm x$ oriented.  (similarly to the adaptive detection
scheme) and the slow increase is due to the slow increase in the
knowledge of $\Omega$ (similarly to direct detection).
As with direct detection, the system state will never
become fully pure. This is due to the double peaks in ${\rm P}(\Omega|\bfi)$,
which insures the $y$ component of the state always averages to zero.

\begin{figure}[t]
\includegraphics[width=0.45\textwidth]{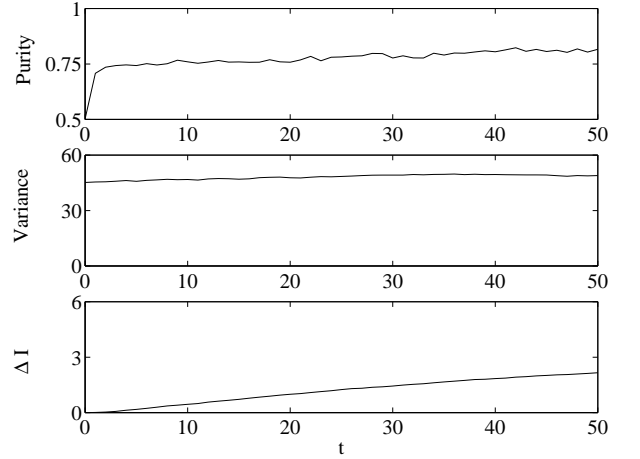}
\caption{\label{fig12} The ensemble average ($n=500$) of the
purity, variance and $\Delta I$ when homodyne $x$ was used. Time
is measured in units of $\gamma^{-1}$.}
\end{figure}

\subsection{Homodyne $y$ Detection} \label{homoy}

Setting the $\Phi$ of the local oscillator to $\pi/2$ allows
measurement of the $y$ quadrature (as $y=\an{-i\sigma+i\sigma\dg}$).
The best estimate states for
the known and unknown $\Omega$ are shown in Fig.~13, for
$\Omega_{\rm true}=5\gamma$. It is seen that
when $\Omega$ is known (solid) this measurement scheme makes the state
coarsely rotate around the Bloch sphere with a purity of one. When $\Omega$ is
unknown (dotted line), Fig.~13 shows that, unlike the previous schemes, the $y$
component does not average to zero. As time
increases the oscillations in the $y$ and $z$ components for the unknown
$\Omega$
case gradually converge to those for the
known $\Omega$ case. This suggests that this scheme can determine
$\Omega_{\rm true}$. This is confirmed by the calculation of ${\rm
P}(\Omega|\bfi)$  shown in Fig.~14. The ability of this scheme to
distinguish  the sign of $\Omega$ can be physically
understood by considering the Bloch representation of \erf{lstoho}.
These stochastic equations are {\em not} invariant
 under the rotation transformation.

\begin{figure}[t]
\includegraphics[width=0.45\textwidth]{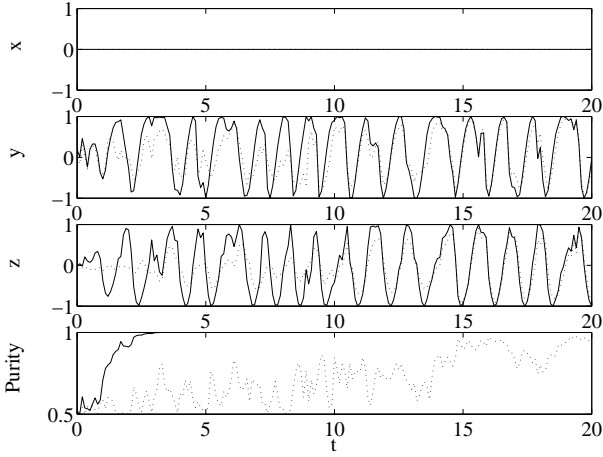}
\caption{\label{fig13} The best estimate states for homodyne $y$
measurement. Details are as in Fig.~2.}
\end{figure}

\begin{figure}[t]
\includegraphics[width=0.45\textwidth]{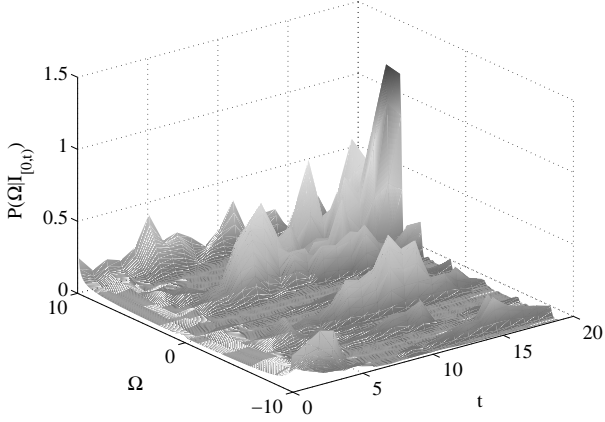}
\caption{\label{fig14} A plot of ${\rm P(\Omega|\bfi)}$ for
homodyne $y$ measurement. Details are as in Fig.~3.}
\end{figure}

 To understand how this scheme reduces the uncertainty
in $\Omega$,
consider the effect for this measurement scheme
\beq
F_{I}dI= \frac{1}{\sqrt{2\pi/dt}} e^{-\half (I-\rt{\gamma}\sigma_{y})^{2}dt}dI.
\eeq
That is, $F_{I}$ is  an unsharp measurement of $y$. Now $y$ is a
variable that is directly affected by $\Omega_{\rm true}$, and
indeed the sign of $y$ reverses if the sign of $\Omega$ reverses.
Even though in each
interval $dt$, $y$ is
measured unsharply, over time this detection scheme will result in a
narrowing of our knowledge of $\Omega$, until infinite time where
it would be fully known. This is further confirmed by the
calculation of the ensemble averages of the three parameters,
 purity, $V$ and $\Delta I$ (Fig.~15).
It is observed that the purity of this state increases up to one, the
$V$ in $\Omega$ reduces substantially in the 50$\gamma^{-1}$ units
of time and $\Delta I$ increases to a value larger than that for
all other schemes.

\begin{figure}[t]
\includegraphics[width=0.45\textwidth]{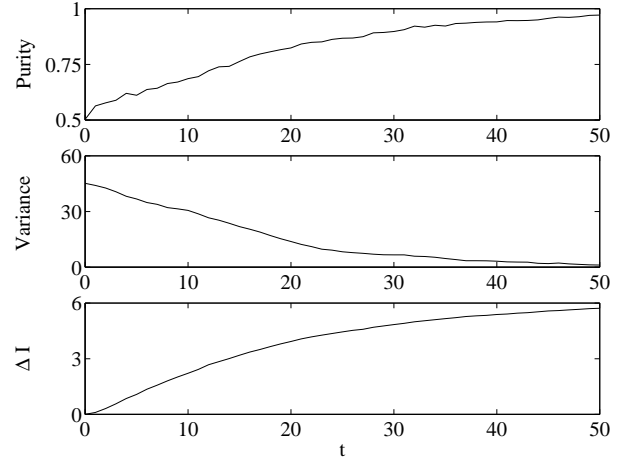}
\caption{\label{fig15} The ensemble average ($n=500$) of the
purity, variance and $\Delta I$ when homodyne detection of the $y$
quadrature was used.}
\end{figure}

\subsection{Heterodyne Detection} \label{hetero}

The last detection scheme considered uses the heterodyne
technique. This detection scheme uses the same arrangement as the
homodyne (see Fig.~9), with the only difference being that the LO is
now detuned from the atom by an amount $\Delta$. This
effectively
results in the LO having a time varying phase of $\Delta t$ with respect
to the driving field. Since the field amplitude is still assumed to be
infinite as in the homodyne case, $\bfi$
will comprises of a string
of real numbers $I$. However, by coarse-graining to obtain the Fourier
components at $\omega=\pm \Delta$, a complex photocurrent is obtained
\cite{WisMil93c}. The continuous set of measurement
operators after the coarse graining approximation ($\Delta dt \gg 1$ but
$\gamma dt \ll 1$) are
\beq
M_{I}=\sqrt{{\Upsilon_{I}}}[1-(i\frac{\Omega}{2}\sigma_{x}+
\frac{\gamma}{2}\sigma\dg\sigma-\rt{\gamma}\sigma
I^{*})dt],
\eeq
where
\beq
{\Upsilon_{I}}d^{2}I
=\frac{dt}{{\pi}}e^{- |I|^{2}dt}d^{2}I.
\eeq
It is easily shown that these measurements operators satisfy the
completeness
condition, \erf{comp}. To do this, one must integrate over the plane
of the complex currents $I$.

As with homodyne, the sample path for $I$ can be obtained from
calculating the probability of getting $I$ in the interval $dt$.
Doing this, one obtains \beq I= [\rt{\gamma}\an{\sigma}+\zeta
(t)], \label{Icurrent} \eeq where $\zeta(t)$ is a complex Gaussian
white noise term, which is formally defined as \cite{Gar85} \bqa
{\rm E}[\zeta (t) \zeta (t')] &=& {\rm E}[\zeta(t)] =0 ,\\
{\rm E}[\zeta^{*}(t)\zeta(t')]&=&\delta(t-t').
\label{noisec}
\eqa

Using the above measurement operators and \erf{Icurrent}, the heterodyne
SME in \ito form is \cite{SME}
\bqa
d\rho_{{\bf I},\Omega}&=&
\rt{\gamma}(\sigma\rho_{{\bf I},\Omega}-\an{\sigma}\rho_{{\bf
I},\Omega})(I^{*}dt
-\rt{\gamma}\an{\sigma\dg}dt) \nl
+\rt{\gamma}(\rho_{{\bf I},\Omega}\sigma\dg-
\an{\sigma\dg}\rho_{{\bf I},\Omega})(Idt-\rt{\gamma}\an{\sigma}dt)\nl
+{\cal L}_{\Omega}\rho_{{\bf I},\Omega}dt.
\label{stohet}
\eqa

For the LSME we again assume that the ostensible probability is that due
just to
the LO, which results in a heterodyne current
$I$ with the same statistics as $\zeta(t)$. With this complex current the
ostensible
probability $\Lambda(I)$ is equal to $\Upsilon_{I}$. This gives a LSME in \ito
form of \cite{SME}
\beq
d\bar\rho_{{\bf I},\Omega}= {\cal L}_{\Omega}\bar\rho_{{\bf I},\Omega}dt
+\rt{\gamma}\sigma\bar\rho_{{\bf I},\Omega}I^{*}dt
+\rt{\gamma}\bar\rho_{{\bf I},\Omega}\sigma\dg Idt.
\label{stochet}
\eeq

Using $\Omega_{\rm true}=5\gamma$, the best estimate state for
known
and unknown $\Omega$ are shown in Fig.~16. It is observed that
for a
known $\Omega$, the state contains attributes of both the homodyne $x$
and $y$ measurement schemes. By this we mean that the state tends to
have a distinct $x$ components, whilst keeping the coarse rotations of the
homodyne $y$ scheme. This is not unexpected as heterodyne is equivalent
to simultaneous homodyne $x$ and $y$ measurements, each of 50$\%$ efficiency
\cite{Lou86}.
 In the unknown $\Omega$ case it is observed that
the $y$ component does not average to zero, suggesting that ${\rm
P(\Omega|\bfi)}$ localizes to $\Omega_{\rm true}$, which is confirmed
by Fig.~17. However, the rate at which ${\rm
P(\Omega|\bfi)}$ converges to $\delta (\Omega -\Omega_{{\rm true}})$ is
much slower
than that of the homodyne $y$ measurement. This is also illustrated in Fig.~18
as the ensemble average $\Delta I$ is not as high. Fig.~18 also
shows the ensemble average of the purity and from this figure it
is seen that it
contains similar properties of both the homodyne $x$ and $y$
schemes. In particular, it has an initial sharp increase, which is due
the state obtaining a large $x$ component (similar to the homodyne $x$
scheme) and as time goes on the purity increases to one due to the
localization of P$(\Omega|\bfi)$ (similar to homodyne $y$).

\begin{figure}[t]
\includegraphics[width=0.45\textwidth]{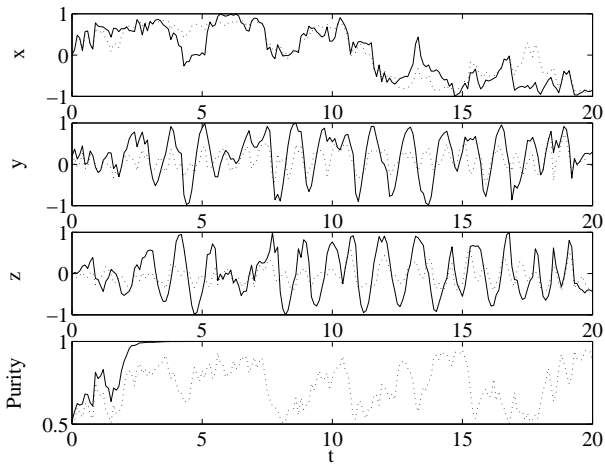}
\caption{\label{fig16}The best estimate states, when heterodyne is
used. Details are as in Fig.~2.}
\end{figure}

\begin{figure}[t]
\includegraphics[width=0.45\textwidth]{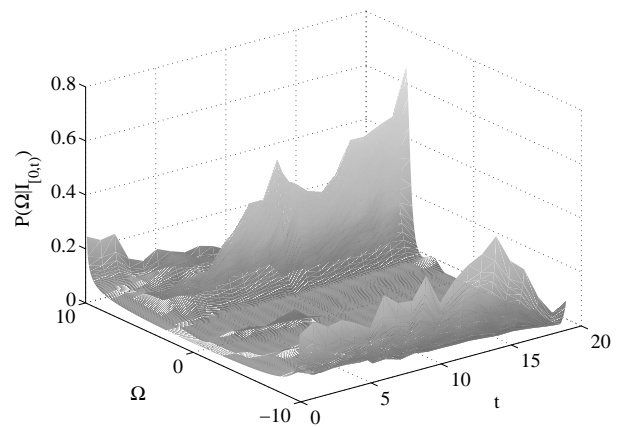}
\caption{\label{fig17} A plot of ${\rm P(\Omega|\bfi)}$ for
heterodyne detection. Details are the same as Fig.~3.}
\end{figure}

\begin{figure}[t]
\includegraphics[width=0.45\textwidth]{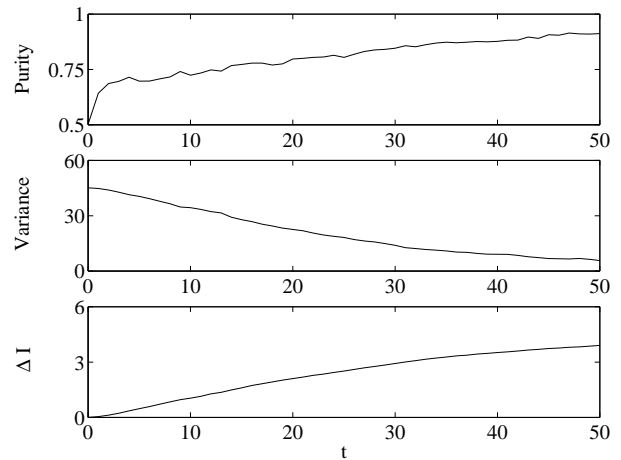}
\caption{\label{fig18} The ensemble average ($n=250$) of the
purity, variance and $\Delta I$ for  heterodyne detection. Time is
measured in units of $\gamma^{-1}$.}
\end{figure}

\section{Discussion}

The results of this paper demonstrate that quantum parameter
and state estimation for a continuously monitored open system
is greatly affected by
the measuring scheme. It was observed that as the measurement time
increased, some detection schemes
had the ability of both reducing our uncertainty in the unknown dynamical
parameter, and producing a conditioned state of high purity,
whereas other schemes could only do one of these, or none (depending
on how the uncertainty in the unknown parameter is quantified).
We re-emphasize that all of the measurement schemes arise from the
same coupling of the system to the environment; all that is different
is how the environment is measured.

The system we considered was a two-level atom with Hamiltonian $\Omega
\sigma_{x}/2$, with spontaneous decay rate $\gamma$. The unknown
dynamical parameter is $\Omega$, the Rabi frequency. We began with
the atom in its
stationary mixed state (depending on $\Omega$) and the prior
distribution of $\Omega$ was that appropriate to an atom at a random
point in a standing
wave with a maximum Rabi frequency $\Omega_{\rm max}=10\gamma$. We
analyzed five different measurement schemes, direct detection, a
particular adaptive scheme \cite{WisToo99}, homodyne detection of the
$x$ quadrature, homodyne of the $y$, and heterodyne. We can summarize
the results of the paper using four different measures of the
effectiveness
of the measurement. The first two relate to the
knowledge obtained about $\Omega$. One is $\Delta I_{l}$, the
long-time ($t >> \gamma^{-1}$) increase in the average information about the
parameter $\Omega$. The other is $V_{l}$, the long-time average variance
in $\Omega$. The next two relate to the knowledge obtained about the
system. One is $p_{l}$, the long-time purity. This measures how much
is known about the system, given the long-time knowledge about the
unknown parameter $\Omega$. The other is $p_{s}$, the short-time
($t =$ a few $\gamma^{-1}$) purity. This time is long enough that,
if $\Omega$ were known, the system would have been more-or-less
completely purified, but short enough that the actual amount of
information obtained about $\Omega$ is small. That is, it measures
how well the measurement can purify the state despite the large
initial uncertainty in the dynamics.

The results of our work is summarized  in the table below, using the
four measures of effectiveness for the five different detection schemes.
Rather than quote figures for these four measures,
we use a rating system ($\star$ to $\star$$\star$$\star$$\star$), the
details of which are explained in the caption. This allows the results
to be taken in at a glance.

\begin{table}[h]
\begin{tabular}{|c|c|c|c|c|c|} 
&\multicolumn{5}{c|}{\bf {Detection Schemes}}\\ \hline
{\bf Measure} & Direct & Adapt & Homo $x$
& Homo $y$ & Hetero \\ \hline
$\Delta I_{l}$ & $\star$$\star$$\star$ & $\star$ & $\star$$\star$
& $\star$$\star$$\star$$\star$ & $\star$$\star$$\star$ \\ \hline
$V_{l}$ & $\star$ & $\star$ & $\star$
& $\star$$\star$$\star$$\star$ & $\star$$\star$$\star$ \\ \hline
$p_{l}$ & $\star$ & $\star$$\star$ & $\star$
& $\star$$\star$$\star$$\star$ & $\star$$\star$$\star$ \\ \hline
$p_{s}$ & $\star$ & $\star$$\star$$\star$$\star$ &
$\star$$\star$$\star$
& $\star$ & $\star$$\star$ \\ 
\end{tabular}
\vspace{0.2cm} \caption{Ratings for the five different detection
schemes, for four different measures. Four $\star$s is the best
rating and one $\star$ the worst. For $\Delta I_{l}$, any rating
above $\star$ indicates that the information about $\Omega$
continues to increase with time, with the lower cut-offs for
$\star$$\star$$\star$ and $\star$$\star$$\star$$\star$ being
$\Delta I_{l} = 2.5$ and $5$ bits respectively at
$t=50\gamma^{-1}$. For $V_{l}$, any rating above $\star$ indicates
a variance in $\Omega$ that decreases, with the upper cut-offs for
$\star$$\star$$\star$ and $\star$$\star$$\star$$\star$ being
$V_{l} = \gamma^{2}$ and $10\gamma^{2}$ respectively at
$t=50\gamma^{-1}$. For $p_{l}$, a rating above $\star$$\star$
indicates a purity that continues to increase with time. For
schemes where the purity saturates, the lower cut-off for
$\star$$\star$ is $p_{l} = 0.9$. For schemes where the purity
continues to increase, the lower cut-off for
$\star$$\star$$\star$$\star$ is $p_{l}=0.95$ at $t=50\gamma^{-1}$.
Finally, for $p_{s}$, the lower cut-offs for $\star$$\star$,
$\star$$\star$$\star$, and $\star$$\star$$\star$$\star$ are,
respectively, $p_{s} = 0.65, 0.75, 0.85$ at $t=3\gamma^{-1}$. In
all cases $\Omega_{\rm max}=10\gamma$.}
\end{table}

From the table it is observed that homodyne $y$ (Sec.~\ref{homoy})
was the best detection scheme by all measures except for the
short-time purification, for which it was the worst. Both of these
aspects are explained by the fact that  this scheme measures
$\sigma_{y}$, the dynamics of which depend strongly on $\Omega$.
Hence the
 measurement record contains a lot of information about $\Omega$,
 including its sign (because rotations over the top of the Bloch
 sphere are different from rotations under the bottom). This also enables
the purity to approach unity as
 time increases. However, for short times, when little information
 about $\Omega$ has been obtained, a $y$ measurement is actually
 very poor for purifying the state. That is because the measurement
 tends to produce states with well-defined values of $y$, and these
 are states that are very sensitive to the rotation around the
 $x$-axis at rate $\Omega$. For a poorly known $\Omega$, this tends to make
 the system state more mixed, so that the purity grows only as the
 information about $\Omega$ increases.

After homodyne $y$ detection, the method that provided most
information about $\Omega$ was direct detection (Sec.~\ref{direct}). Under
direct
detection, the count rate is proportional to $\sigma_{z}+1$, and
(like $\sigma_{y}$), the dynamics of $\sigma_{z}$ depend strongly upon
$\Omega$, due to the Rabi rotations around the $x$-axis. However, in
terms of $\sigma_{z}$, rotations around the $+x$-axis from the ground state
are
indistinguishable from rotations around the $-x$-axis. Hence the
measurement cannot distinguish the sign of $\Omega$ and there is no
change in the ensemble averaged variance as time increases. As a
consequence, the purity saturates at a low value. The short time
purification is poor also, for a similar reason to that for homodyne
$y$ detection.

The adaptive detection is almost complementary in its qualities to
homodyne $y$ detection. As explained in Sec.~\ref{adapt}, it yields
almost no information about $\Omega$, because the rate of detections
in steady state is independent of $\Omega$. In particular, it yields
no information about the sign of $\Omega$, so the variance is constant.
As a consequence, the purity does not approach unity. Nevertheless, it
does approach a quite high value, of over $1-\gamma/(\rt{2}\Omega_{\rm
max})$, which is $0.93$ for $\Omega_{\rm max}=10\gamma$. This is because the
conditioned states are, for large $\Omega$, asymptotically
independent of $\Omega$, as they approach $\sigma_{x}$
eigenstates. This explains why the adaptive scheme gives the best
results for short-time purification: the conditioned states are
almost unaffected by the uncertainty in $\Omega$.

Homodyne $x$ detection (Sec.~\ref{homox}) is in many ways similar to the
adaptive
scheme, and this is readily understandable since it would be expected to
produce conditioned states tending towards $\sigma_{x}$ eigenstates.
Like adaptive (and direct) detection, the sign of $\Omega$ is
indeterminable so the variance is constant. Hence the final purity does not
approach unity. Although its asymptotic value is
not as high as that for adaptive detection, it is higher than that
for direct detection. This is as expected, since the  conditioned
states, being imperfectly localized towards the $x$-eigenstates,
 are still affected by $\Omega$. This also explains why the initial
 purification is not quite as good as for adaptive detection, and why
 information continues to be gained (albeit slowly) as time increases.

The final scheme, heterodyne detection (Sec.~\ref{hetero}), is most
easily understood by viewing it as an equal mixture of homodyne $x$ and
homodyne $y$ detection, which is in fact a completely rigorous
viewpoint. All of the ratings for heterodyne detection
are intermediate between those for the two homodyne schemes.

In conclusion, we have shown that gaining knowledge about an unknown dynamical
parameter by monitoring the system is a quite different phenomenon
from gaining knowledge
about the system itself. We have also distinguished different sorts
of knowledge acquisition with distinct characteristics: for the unknown
parameter, information gain (in bits) versus reducing the variance; and
for the system, short-time purity gain versus long-time purity gain.
The ability to acquire knowledge in these various ways is extremely
sensitive to
the choice of monitoring scheme (which does
not affect the average evolution of the system). For the system we
investigated, explaining the particulars of this sensitivity depends
upon a detailed understanding of the conditional dynamics of the
system. Our discoveries may have important implications for the
suitability of different  quantum
feedback-control techniques \cite{Wis95b,Doh00} in experimental
systems with unknown dynamical parameters.
Another direction for future
work could be to investigate the effect of realistic
imperfections in the detection schemes on state and parameter estimation
in open quantum systems.

\acknowledgments

HMW would like to acknowledge formative
conversations with Andrew Doherty.


\begin{thebibliography}{}
\bibitem{perscom}
The same idea has been considered by A. Doherty (pers. comm.).
\bibitem{SME}
These equations can be derived
 by considering photodetection and taking the limit as
the local oscillator becomes infinite \cite{Car93b,WisMil93a}, or,
more directly, by projection of the environment into
eigenstates of the field quadrature \cite{GoeGra94b}. Heterodyne
detection can also be treated in the first way \cite{WisMil93c} or by
projection onto eigenstates of the field annihilation
operator \cite{WisTh}.
\end{thebibliography}
\end{document}